\documentclass[10pt]{article}

\usepackage{amsmath}

\usepackage{array}

\usepackage{appendix}

\usepackage{tocloft}                   

\usepackage{graphicx}

\usepackage{amsfonts}

\usepackage{amssymb}

\usepackage{mathrsfs}

\usepackage{yfonts}

\usepackage{euscript}

\usepackage{centernot}                 

\usepackage{ifsym}                     

\usepackage{upgreek}

\usepackage{mathtools}

\usepackage{bm}

\usepackage{color}

\usepackage{slantsc}
\usepackage{calligra}

\usepackage{bbold}          

\usepackage[T1]{fontenc}

\usepackage{epsf}

\usepackage{latexsym}

\usepackage{tipa}

\usepackage{makeidx}

\makeindex

\def\be{\begin{equation}}

\def\ee{\end{equation}}

\def\beq{\begin{equation}}

\def\eeq{\end{equation}}

\def\bea{\begin{eqnarray}}

\def\eea{\end{eqnarray}}

\def\!{\hspace{-1.6667em}}

\def\m{\mbox{ }}

\def\mma {\m , \m \m }

\def\!{\hspace{-1.6667em}}

\def\n{\noindent}

\def\u{\underline}

\def\es{\m = \m}

\def\:={\m := \m}

\def\=:{\m =: \m}

\def\biq{\mbox{\boldmath$q$}}

\def\sbiQ{\mbox{\scriptsize\boldmath$Q$}}

\def\bnabla{\mbox{\boldmath$\nabla$}}               

\def\miu{\mbox{$u$}}

\def\miv{\mbox{$v$}}

\def\momega{\mbox{$\omega$}}

\def\miw{\mbox{$w$}}

\def\mix{\mbox{$x$}}

\def\miy{\mbox{$y$}}

\def\miz{\mbox{$z$}}

\def\miq{\mbox{$q$}}

\def\barr{\bar{r}}

\def\sumi2{\sum\mbox{}_{\mbox{}_{\mbox{\scriptsize $i$=1}}}^2}

\def\sumi3{\sum\mbox{}_{\mbox{}_{\mbox{\scriptsize $i$=1}}}^3}

\def\sumABcycles3{\sum\mbox{}_{\mbox{}_{\mbox{\scriptsize cycles $A,B$=1}}}^{3}}

\def\sumCDcycles3{\sum\mbox{}_{\mbox{}_{\mbox{\scriptsize cycles $C,D$=1}}}^{3}}

\def\sumj3{\sum\mbox{}_{\mbox{}_{\mbox{\scriptsize $j$=1}}}^3}

\def\sumk3{\sum\mbox{}_{\mbox{}_{\mbox{\scriptsize $k$=1}}}^3}

\def\prodiA1{\prod\mbox{}_{\mbox{}_{\mbox{\scriptsize $i$=1}}}^{A - 1}}

\def\d{\textrm{d}}                                                  

\def\pa{\partial}                                                   

\def\sbiQ{\mbox{\scriptsize \boldmath$Q$}}

\def\Hilb{\mbox{{\boldmath$\mathfrak{H}$}ilb}}                 

\def\sii{\mbox{\scriptsize$i$}}

\def\siN{\mbox{\scriptsize$N$}}

\def\sin{\mbox{\scriptsize$n$}}

\def\Phase{\mbox{{\boldmath$\mathfrak{P}$}hase}}                     

\def\bFrR{\mbox{\boldmath$\mathfrak{R}$}}                            
\def\Rig-Phase{\bFrR\mbox{ig-}\Phase}                                

\def\Positive-Modespace{\mbox{{\boldmath$\mathfrak{M}$}odespace$^+$}}

\def\POSITIVE-MODESPACE{\mbox{{\boldmath$\mathfrak{M}$}ODESPACE$^+$}}

\def\Kin-Hilb{\mbox{{\boldmath$\mathfrak{K}$}in-\Hilb}}                     

\def\Mid-Hilb{\mbox{{\boldmath$\mathfrak{M}$}id-\Hilb}}                     

\def\Dyn-Hilb{\mbox{{\boldmath$\mathfrak{D}$}yn-\Hilb}}                     

\def\5Star{\mbox{\Large$\star$}}              

\begin{document}

\begin{titlepage}

\begin{center}

\vspace{0.1in}

\Large{\bf Specific PDEs for Preserved Quantities in Geometry. III.} 

\vspace{0.1in}

\Large{\bf 1-d Projective Transformations and Subgroups} \normalsize

\vspace{0.1in}

{\large \bf Edward Anderson$^*$}

\vspace{0.1in}

\end{center}

\begin{abstract}

We extend finding geometrically-significant preserved quantities by solving specific PDEs to 1-$d$ projective transformations and subgroups.
This can be viewed not only as a purely geometrical problem but also as a subcase of finding physical observables, 
and furthermore as part of extending the comparative study of Background Independence level-by-level in mathematical structure to include projective structure.   
Full 1-$d$ projective invariants are well-known to be cross-ratios.  
We moreover rederive this fact as the unique solution of 1-$d$ projective geometry's preserved equation PDE system. 
We also provide the preserved quantities for the 1-$d$ geometries whose only transformations are 1) special-projective transformations $Q$, giving differences of reciprocals. 
2) $Q$ alongside dilations $D$, now giving ratios of difference of reciprocals.
This analysis moreover firstly points to a new interpretation of cross-ratio: those ratios of differences that are concurrently differences of reciprocals, 
and secondly motivates 1) and 2) as corresponding to bona fide and distinctive Geometries.  
	
\end{abstract}

\n Mathematics keywords: Geometrical automorphism groups and the corresponding preserved quantities.  
Differential-geometric PDEs. Projective Geometry. Shape Theory. Foundations of Geometry.  

\m

\n PACS: 04.20.Cv, 04.20.Fy, Physics keywords: observables, Background Independence.

\m

\vspace{0.1in}
  
\n $^*$ Dr.E.Anderson.Maths.Physics@protonmail.com

\section{Introduction}

\n We continue our program of geometrical preserved quantities (in Article I \cite{PE-1}'s sense) being 
systematically derived as solutions to PDE systems -- preserved equations -- treated as Free \cite{CH1} Characteristic Problems \cite{CH2, John}.  
\cite{PE-1} moreover provides specific methods of solution for these PDE systems. 
Articles I and II having considered this for Similarity and Affine Geometry respectively, 
we now turn to the case of 1-$d$ Projective Geometry \cite{Desargues, VY10, HC32, S04, Stillwell, Hartshorne}, 
Projective Geometry playing a notable role in the Foundations of Geometry \cite{Hilb-Ax, HC32, Coxeter, Stillwell, 8-Pillars}
The corresponding geometrical automorphism group is 
\be 
Proj(1)  \es  PGL(2, \mathbb{R})  \m , 
\ee
`PGL' standing for projective general linear group.  
This has three generators: a translation $P$, a dilation $D$ and a {\it special-projective transformation} $Q$ 
(c.f.\ the more commonly used concept of a special-conformal transformation and generator). 

\m 

\n $P$ and $D$ already being the 1-$d$ similarity group $Sim(1)$'s generators, they were already considered in Article I. 
We thus begin by considering $Q$ by itself in Secs 2 and 3. 
While the ensuing automorphism group is isomorphic to that of the 1-$d$ translation $P$, $Q$ nonetheless possesses a distinct notion of preserved quantity: 
{\it differences of reciprocals} rather than just differences.
Upon including $D$ as well as $Q$ (Sec 4), the automorphism group deviates from the dilatations ($P$ and $D$) by just a sign, 
and yet again a distinct notion of preserved quantity ensues: {\it ratios of differences of reciprocals} rather than just ratios of differences.
We take these distinctive geometrical invariants to further motivate study of these `Iso-Translational' and `Para-Dilatational' Geometries, 
which moreover in 1-$d$ further coincide with `Iso-Euclidean' and `Iso-Similarity' = `Iso-Affine' Geometries respectively.

\m 

\n Full 1-$d$ projective invariants are well-known to be {\it cross-ratios}, 
quantities whose invariance properties can already be inferred from the work of Pappus \cite{Pappus}, 
and whose modern-era development started with the elder Carnot \cite{Carnot}, for all that the name `cross-ratio' itself was only coined in subsequent work by Clifford \cite{Clifford}.  
We moreover derive that 1-$d$ projective preserved quantities are suitably smooth functions of cross-ratios in Secs 5 and 6, 
establishing these to be the {\sl unique} functional form solving the 1-$d$ projective preserved equations system's Free Characteristic Problem in Secs 7 and 8.   
The current Article's analysis points moreover to a new interpretation of cross-ratio: 
cross-ratio functional dependence is that functional dependence which is concurrently of ratios of differences and of differences of reciprocals. 
With $P$ and $Q$ being inconsistent by themselves, this completes the study of the geometrically-significant continuous subgroups of $Proj(1)$.  

\m 

\n Preserved quantities as conceived of in the current Series of Articles are moreover underlied by consideration of constellations, constellation spaces, shapes and 
shape spaces \cite{Kendall84, Kendall89, Small, Kendall, FileR-Quad-I, PE16, ABook, I-II-III-Minimal-N}. 
In the context of Projective Geometry, the corresponding Projective Shape Theory has been developed and reviewed in particular in \cite{MP05, Bhatta, PE16, KKH16}.   
Its main application to date is to Shape Statistics \cite{Kendall, JM00, Bhatta, DM16, PE16} in connection with 

\end{titlepage}

\n Image Analysis \cite{Images} and Computer Vision \cite{CV}.  
Projective Geometry has largely not yet entered comparative study \cite{I89-I91, ABook, ASoS, AMech, Minimal-N-2} 
of Background Independence \cite{DiracObs, BI, BI-2, Giu06, ABeables, AObs2, AObs3, AObs4, APoT, ABook, 5-6-7} in Foundational and Theoretical Physics. 
Upcoming preprints in this regard will be linked here to subsequent versions of the current preprint as regards this substantial research frontier.

\section{Special-projective transformations and preserved equations}

In $d$-dimensional Projective Geometry, the special-projective transformation's generator is 
\be 
Q^a  :=  x^a x^b \pa_b                                                                          \m . 
\ee 
The special-projective preserved equation is thus  

\n\be 
\sum_{I = 1}^N Q^a(\u{q}^I) \sbiQ  \es  \sum_{I = 1}^N q^{aI} q^{bI} \pa_{q^{bI}} \sbiQ  \es  0   \m .  
\ee
Special-projective transformations close by themselves, forming the geometrical -- if hitherto nonstandard -- automorphism group 
\be 
P\mbox{-}Iso\mbox{-}Tr(d)                                                                       \m .
\ee
`Iso' stands here for isomorphic, and the $P$ prefix for projective version, there also being a conformal version denoted by a $C$ prefix in Article V. 

\m

\n While this is isomorphic to the non-compact $d$-dimensional Abelian group of translations, 
\be 
P\mbox{-}Iso\mbox{-}Tr(d)  \m \cong \m  \mathbb{R}^d  \m \cong \m  Tr(d)                        \m ,
\ee 
it clearly involves a different representation of generators, $x^a x^b \pa_b$ rather than the $\pa_b$ used for the translations.  
We show in the current Article that this furthermore leads to $P\mbox{-}Iso\mbox{-}Tr(d)$ having different geometrical preserved quantities than $Tr(d)$, 
strengthening the position that $P\mbox{-}Iso\mbox{-}Tr(d)$ indeed corresponds to a distinct Geometry in its own right.  

\m 
 
\n In 1-$d$, the special-projective generator moreover simplifies to 
\be 
Q  \es  x^2 \, \frac{\d}{\d x}                                                                   \m . 
\ee 
This is one of the ways in which 1-$d$ is a distinguished special case.  
The special-projective preserved equation is now 

\n\be 
\sum_{I = 1}^N q^{I \, 2} \pa_{q^{I}} \sbiQ  \es  0       \m .  
\label{SP-PE}
\ee
Moreover, in 1-$d$, 
\be 
\mathbb{R}  \es  P\mbox{-}Iso\mbox{-}Tr(1)  
            \es  P\mbox{-}Iso\mbox{-}Eucl(1)                                                     \m ,  
\ee 
due to the absence of rotations and of nontrivial special-linear transformations respectively.  
This is a second way in which 1-$d$ is a distinguished case.

\section{Piecemeal solution of the 1-$d$ special-projective preserved equation}

For $N = 1$, using the notation $q_1 = x$, our preserved equation reduces by the flow method to the ODE 
\be 
x^2 \, \frac{\d \sbiQ}{\d x} = 0        \m .  
\ee 
So for $x \neq 0$, this reduces to 
\be 
\frac{\d \sbiQ}{\d x} = 0               \m , 
\ee 
and thus admits just the trivial solution, 
\be 
\sbiQ = const \m .
\ee 
For $x = 0$, $\sbiQ$ is a free function, but as for the dilational case in Article 1, this entails having no freedom in moving away from $x = 0$.  

\m 

\n $N = 2$ is minimal as regards having a nontrivial solution.
Using also the notation $q_2 = y$, our preserved equation is now the PDE 
\be 
( x^2 \pa_x + y^2 \pa_y ) \sbiQ = 0         \m . 
\label{Proj1N2}
\ee 
Being a homogeneous-linear PDE in 2 variables, this is equivalent to the ODE 
\be 
\frac{\d x}{x^2}  \es  \frac{\d y}{y^2} 
\ee 
which is amenable to direct integration, giving  
\be 
- \frac{1}{x} \es - \frac{1}{y} + u      \m , 
\ee 
i.e.\ 
\be 
u  \es  \frac{1}{x} - \frac{1}{y}      \m . 
\label{Char-N2}
\ee 
By the chain-rule, any 
\be 
\sbiQ  \es  \sbiQ(u)  \es \sbiQ\left(  \frac{1}{x} - \frac{1}{y}  \right) 
\label{PP-N2}
\ee 
moreover also solves:  
\be
(x^2\pa_x + y^2 \pa_y)\sbiQ(u)  \es  \sbiQ^{\prime}(u)(x^2\pa_x + y^2 \pa_y)\left(  \frac{1}{x} - \frac{1}{y}  \right)  
                               \es  \sbiQ^{\prime}(u) \left(  - \frac{x^2}{x^2} + \frac{y^2}{y^2}  \right)
                               \es  \sbiQ^{\prime}(u)(-1 + 1) 
                                =   0                          \m , 							   
\ee
where $\mbox{}^{\prime}  :=  \d/\d u$. 
Alternatively, by the flow method, PDE (\ref{Proj1N2}) is equivalent to the ODE system
\be 
\dot{x}    = x^2  \m , 
\ee 
\be 
\dot{y}    = y^2  \m , 
\ee 
\be 
\dot{\sbiQ} = 0    \m , 
\ee 
to be treated as a Free Characteristic Problem. 
Integrating, 
\be 
t  \es   - \frac{1}{x} + u  \m ,
\label{ODE-1}
\ee 
\be 
t  \es   - \frac{1}{y}      \m ,
\label{ODE-2}
\ee 
\be 
\sbiQ  = \sbiQ(u)  \m . 
\label{ODE-3}
\ee 
Next, eliminating $t$ between (\ref{ODE-1}-\ref{ODE-2}), we obtain the form of the characteristic coordinate (\ref{Char-N2}). 
Finally substituting this in (\ref{ODE-3}), we recover the form (\ref{PP-N2}) for the preserved quantities.  

\m 

\n Extending to the arbitrary-$N$ case, the preserved equation PDE (\ref{SP-PE}) is equivalent by the flow method to the ODE system
\be 
\dot{q}^I    = q^{I \, 2}  \m , 
\ee 
\be 
\dot{\sbiQ} = 0             \m , 
\ee 
to be treated as a Free Characteristic Problem. 
Integrating and splitting off $q^N$ for distinct treatment, 
\be 
t  \es   - \frac{1}{q^i} + u_i  \m ,
\label{ODE-13}
\ee 
\be 
t  \es   - \frac{1}{q^N}        \m ,
\label{ODE-14}
\ee 
\be 
\sbiQ  = \sbiQ(u^i)  \m . 
\label{ODE-15}
\ee 
Next, eliminating $t$ from (\ref{ODE-14}) in (\ref{ODE-13}), we obtain the form of the characteristic coordinates,
\be 
u^i  \es  \frac{1}{q^i} - \frac{1}{q^N}      \m ,  
\label{Char-N}
\ee 
Finally substituting these in (\ref{ODE-15}), we obtain the functional form 
\be 
\sbiQ  \es  \sbiQ\left( \frac{1}{q^i} - \frac{1}{q^N}   \right) 
\ee 
for the preserved quantities. 
We finally summarize 1-$d$ special-projective preserved quantities by 
\be 
\sbiQ \es \sbiQ(\, {\bm{/-/}} \,)  \m :
\ee 
suitably-smooth functions of differences of reciprocals, for which we have introduced the shorthand ${\bm{/-/}}$. 

\m 

\n Subsequent restrictions of note render $N = 3$ and 4 as minimal cases of interest as well. 
For $N = 3$, using $q_3 := z$ as well, 
\be 
\sbiQ  \es  \sbiQ(u, \, v)  \:=  \sbiQ\left( \frac{1}{x} - \frac{1}{z} \mma  \frac{1}{y} - \frac{1}{z}   \right)  \m . 
\ee 
\n For $N = 4$, using $q_4 =: w$ as well, 
\be 
\sbiQ  \es  \sbiQ(u, \, v, \, \omega) \:=  \sbiQ\left( \frac{1}{w} - \frac{1}{z} \mma  \frac{1}{x} - \frac{1}{z} \mma  \frac{1}{y} - \frac{1}{z}   \right)  \m . 
\ee

\section{$P\mbox{-}Para\mbox{-}Dilatat(1)$}

The $Q$ and $D$ generators mutually close to form the geometrical automorphism group 
\be 
P\mbox{-}Para\mbox{-}Dilatat(1)  \es  \mathbb{R} \rtimes \mathbb{R}_+   \es  P\mbox{-}Para\mbox{-}Sim(1)  \es  P\mbox{-}Para\mbox{-}Aff(1)        \m .
\ee 
`Para' refers to this case {\sl not} coinciding isomorphically with 
\be 
Dilatat(1)  \es  \mathbb{R} \rtimes \mathbb{R}_+  \es  Sim(1)  \es  Aff(1)  \m . 
\ee 
It is possible for two non-isomorphic groups to both be of the form $A \rtimes B$ 
because the semidirect product operation is {\sl not} precisely defined 
unless one supplements it by identifying which map is involved \cite{Cohn}.
On the present occasion, this non-isomorphism is clear from 
\be 
\mbox{\bf [} Q \mbox{\bf ,} \, D \mbox{\bf ]} \m \sim \m  Q \m \mbox{ and } \m \mbox{\bf [} P \mbox{\bf ,} \, D \mbox{\bf ]} \m \sim \m  P 
\ee  
differing by a sign in their more detailed right-hand-sides. 
We shall encounter further $P$- and $C$- $Para$ groups of geometrical automorphisms in Articles IV and V.  

\m 

\n The corresponding preserved equations are a system of 2 equations, 

\n\be 
\biq \circ \bnabla \sbiQ  \es  \sum_{I = 1}^N q^I \pa_I \sbiQ  \es  0         \m , 
\label{Sis-1}
\ee 

\n\be 
\sum_{I = 1}^N q^{I \, 2} \pa_I \sbiQ  \es  0                               \m . 
\label{Sis-2}
\ee 
Counting out, $N = 3$ is now minimal so as to realize nontrivial preserved quantities.
In this case, our system reduces to 
\be 
( x \, \pa_x + y \, \pa_y + z \, \pa_z ) \sbiQ  =  0                                  \m , 
\label{Dil1N3}
\ee 
\be 
( x^2 \pa_x + y^2 \pa_y + z^2 \pa_z ) \sbiQ  =  0                                     \m . 
\label{Proj1N3-2}
\ee 
But we solved these equations piecemeal in Secs I.8 and III.3, so we have the {\it compatibility equation} 
\be 
\sbiQ\left( \frac{x}{z} \mma \frac{y}{z} \right)  \es  \sbiQ\left(  \frac{1}{x} - \frac{1}{z} \mma  \frac{1}{y} - \frac{1}{z}  \right)  \m .
\ee 
{\bf Lemma 1} This is solved by 
\be 
\sbiQ\left(  \frac{\frac{1}{\mix} - \frac{1}{\miz}}{\frac{1}{\miy} - \frac{1}{\miz}}  \right)  \m . 
\ee 
{\u{Derivation}}.
On the one hand, this is manifestly a function of 
\be 
\frac{1}{x} - \frac{1}{z} \mma  \frac{1}{y} - \frac{1}{z} \m \mbox{ alone}                    \m .
\ee
On the other hand, 
\be 
\frac{  \frac{1}{\mix} - \frac{1}{\miz}  }{  \frac{1}{\miy} - \frac{1}{\miz}  }  \es   
\frac{  \frac{1}{\miz} \left( \frac{\miz}{\mix} - 1 \right)  }{  \frac{1}{\miz} \left( \frac{\miz}{\miy} - 1 \right)  }  
                                                             \es   \frac{                   \frac{\miz}{\mix} - 1   }{                          \frac{\miz}{\miy} - 1  }
															 \es   \frac{             \left( \frac{\mix}{\miz} \right)^{-1} - 1  }{  \left( \frac{\miz}{\miy} \right)^{-1} - 1  } \m , 
\ee 
which is manifestly a function of 
\be 
\frac{x}{z}  \mma  \frac{y}{z}  \m  \mbox{ alone}            \m . \m \m  \Box 
\ee 
Note also the following alternative form for this answer,  
\be 
\sbiQ  \es  \sbiQ\left( \frac{y(z - x)}{x(z - y)} \right)      \m . 
\ee  
$N = 4$ is also of interest, as the minimal case in the next section. 
In this case, our system reduces to 
\be 
( w \pa_w + x \pa_x + y \pa_y + z \pa_z ) \sbiQ = 0       \m , 
\label{Dil1N4}
\ee 
\be 
( w^2 \pa_w + x^2 \pa_x + y^2 \pa_y + z^2 \pa_z ) \sbiQ = 0     \m . 
\label{Proj1N4-2}
\ee 
But we solved these equations piecemeal in Secs I.7 and III.2, so we have the {\it compatibility equation} 
\be 
\sbiQ\left( \frac{w}{z} \mma \frac{x}{z} \mma \frac{y}{z} \right)  \es  
\sbiQ\left(  \frac{1}{w} - \frac{1}{z} \mma  \frac{1}{x} - \frac{1}{z} \mma \frac{1}{y} - \frac{1}{z}  \right)                          \m .
\ee 
This is solved similarly by 
\be 
\sbiQ  \es  \sbiQ\left(  \frac{\frac{1}{\miw} - \frac{1}{\miz}}{\frac{1}{\miy} - \frac{1}{\miz}} \mma  \frac{\frac{1}{\mix} - \frac{1}{\miz}}{\frac{1}{\miy} - \frac{1}{\miz}}  \right)  
      \es  \sbiQ\left( \frac{y(z - w)}{w(z - y)} \mma \frac{y(z - x)}{x(z - y)} \right)                                                 \m .  
\ee 	 
Finally, in the arbitrary-$N$ case, (\ref{Sis-1}, \ref{Sis-2}) lead to the compatibility equation 
\be 
\sbiQ\left( \frac{\miq^{\sii}}{\miq^{\siN}} \right)  \es  
\sbiQ\left(  \frac{1}{\miq^{\sii}} - \frac{1}{\miq^{\siN}}  \right)                                                                     \m ,
\ee 
which is solved by 
\be 
\sbiQ  \es  \sbiQ\left(  \frac{\frac{1}{\miq^{\barr}} - \frac{1}{\miq^{\siN}}}{\frac{1}{\miq^{\sin}} - \frac{1}{\miq^{\siN}}}  \right)   
      \es  \sbiQ\left( \frac{q^n(q^N - q^{\barr})}{q^{\barr}(q^N - q^n)} \right)
	  \es  \sbiQ\left( \, {\bm{/-/}} \, \bm{\mbox{\Large /}}  \, {\bm{/-/}} \, \right)                                                      \m :   
\ee
suitably-smooth functions of ratios of differences of reciprocals.  

\m 

\n{\bf Remark 1} These are geometrical preserved quantities, and distinctively different from those of any of the hitherto well-studied Geometries.

\section{$Proj(1)$ system of preserved equations}

\n First note that considering translations $P$ alongside special-projective transformations $Q$ and no other generators is inconsistent by the integrability relation   
\be 
\mbox{\bf [} P \mbox{\bf ,} \, Q \mbox{\bf ]} \m \sim \m D   \m .
\ee
forces the dilation generator $D$ to be included as well.
Taking all three of these generators together, one has the group of 1-$d$ projective transformations,  
\be 
Proj(1)  \es  PGL(2, \mathbb{R})  \m . 
\ee 
\n The corresponding system of preserved equations is   

\n\be 
\sum_{I = 1}^N \pa_I \sbiQ  \es  0                  \m , 
\ee 

\n\be 
\biq \circ \bnabla \sbiQ  \es  \sum_{I = 1}^N q^I\pa_I \sbiQ  \es 0  \m , 
\ee 

\n\be 
\sum_{I = 1}^N q^{I \, 2}\pa_I \sbiQ \es 0                               \m . 
\ee 
Counting out, $N = 4$ is now minimal to support nontrivial solutions. 

\m

\n The sequential working based on passing to centre of mass coordinates moreover also fails, 
as a consequence of this integrability involving $P$ more intimately in $Proj(1)$'s Lie algebra 
                                                          than a simple semidirect product addendum.

\section{$Proj(1)$ preserved quantities for $N = 4$}

\n However, for $N = 4$, however solved the first pair of these equations in Sec I.8 and the last equation in Sec III.3, so we have the compatibility equation 
\be 
\sbiQ\left(  \frac{1}{w} - \frac{1}{z} \mma  \frac{1}{x} - \frac{1}{z} \mma \frac{1}{y} - \frac{1}{z}  \right)  \es 
\sbiQ\left(  \frac{w - z}{y - z} \mma \frac{x - z}{y - z}  \right)                                                                                                       \m . 
\ee
\n {\bf Lemma 2} This is solved by 
\be 
\sbiQ  \es  \sbiQ\left(\frac{(w - z)(x - y)}{(w - y)(x - z)}\right)  
      \:=  \sbiQ\left( \, [z, \, y; \, w, \, x] \, \right)                                                                                                               \m ,
\ee 
for $[ \m , \m ; \m , \m ]$ the quaternary {\it cross-ratio} operation. 

\m 

\n{\u{Derivation}} On the one hand, 
\be
\frac{(w - z)(x - y)}{(w - y)(x - z)}  \es  \frac{\frac{(\miw - \miz)(\mix - \miy)}{\miw\mix\miy\miz}}  {\frac{(\miw - \miy)(\mix - \miz)}{\miw\mix\miy\miz}} 
                                       \es  \frac{  \frac{\miw - \miz}{\miw\miz} \times \frac{\mix - \miy}{\mix\miy}   }{  \frac{\miw - \miy}{\miw\miy} \times  \frac{\mix - \miz}{\mix\miz}  }
                                       \es  \frac{  \left( \frac{1}{\miz} - \frac{1}{\miw} \right)\left( \frac{1}{\miy} - \frac{1}{\mix} \right)  }
                                                 {  \left( \frac{1}{\miy} - \frac{1}{\miw} \right)\left( \frac{1}{\mix} - \frac{1}{\miz} \right)  }
                                       \es  \frac{  -\left( \frac{1}{\miw} - \frac{1}{\miz} \right)
									                 \left( \left( \frac{1}{\miy} - \frac{1}{\miz} \right) - \left( \frac{1}{\mix} - \frac{1}{\miz} \right) \right)  }
                                                 {   \left( \left( \frac{1}{\miy} - \frac{1}{\miz} \right) - \left( \frac{1}{\miw} - \frac{1}{\miz} \right) \right)
												     \left( \frac{1}{\mix} - \frac{1}{\miz} \right)  }                                                                  \m ,
\ee 
which is manifestly a function of 
\be 
\frac{\frac{1}{\miw} - \frac{1}{\miz}}{\frac{1}{\miy} - \frac{1}{\miz}} \mma  \frac{\frac{1}{\mix} - \frac{1}{\miz}}{\frac{1}{\miy} - \frac{1}{\miz}} \m \mbox{ alone}  \m .
\ee 
On the other hand, 
\be  
\frac{  (w - z)(x - y)}  {  (w - y)(x - z)  }  \es  \frac{  (w - z)((x - z) - (y - z))  }{  ((w - z) - (y - z))(x - z)  }  
                                               \es  \frac{  \frac{\mix - \miz}{\miy - \miz}  - 1  }{  \frac{\mix - \miz}{\miy - \miz}  }           \times 
											        \frac{  \frac{\miw - \miz}{\miy - \miz}  }{  \frac{\miw - \miz}{\miy - \miz} - 1  }                                 \m , 
\ee 
which is manifestly a function of 
\be 
\frac{w - z}{y - z} \mma \frac{x - z}{y - z} \m \mbox{ alone} \m . \m \m \Box 
\ee 
\n{\bf Remark 1} We can now {\sl characterize cross-ratios' functional dependence as being concurrently of ratios of differences and of differences of reciprocals}, 
\be 
{\bm{;}}  \es  {\bm{ - / - }} \m \bigcap \m {\bm{ / - / }}  \m . 
\ee 
This characterization is enlightening since it is in terms of simpler, more well-known and more intuitive operations.

\section{Uniqueness of cross-ratios in 1-$d$. I. $N = 4$}

The above type of working is still open to the possibility that 
\be 
\frac{(a - b)(c - d)(e - f) ...}{(\sigma(a) - \sigma(b))(\sigma(c) - \sigma(d))(\sigma(e) - \sigma(f))...}
\ee 
for $\sigma$ a permutation, provides further independent functions of both concurrently  differences of reciproals and ratios of differences. 
We now dismiss this possibility by use of the sequential chain rule method.

\m 

\n We first present this for $N = 4$.  
We begin by solving the special-projective preserved equation piecemeal, as per Sec 3.
Then by the chain rule, 
\be 
\pa_w = u_w \pa_w + v_w \pa_v + \omega_{w} \pa_{\omega}  \es  - \frac{1}{w^2} \pa_u                       \m , 
\ee 
\be 
\pa_x = u_x \pa_w + v_x \pa_v + \omega_{x} \pa_{\omega}  \es  - \frac{1}{x^2} \pa_v                       \m , 
\ee 
\be 
\pa_y = u_w \pa_y + v_w \pa_v + \omega_{y} \pa_{\omega}  \es  - \frac{1}{y^2} \pa_{\omega}                       \m , 
\ee 
\be 
\pa_z = u_z \pa_w + v_z \pa_v + \omega_{z} \pa_{\omega}  \es  \frac{1}{z^2} \, ( \pa_u + \pa_v + \pa_{\omega} )  \m . 
\ee 
\n The dilational preserved equation then becomes 
\be 
0  \es  \left( \left(\frac{1}{w} - \frac{1}{z}   \right) \pa_u + \left( \frac{1}{x} - \frac{1}{z}  \right) \pa_v  + \left( \frac{1}{y} - \frac{1}{z}  \right) \pa_{\omega}\right) \sbiQ 
   \es       ( u \, \pa_u  +  v \, \pa_v  + w \, \pa_{\omega} ) \sbiQ                                              \m .  
\ee 
As this has the same form as the original dilational equation but for one object less, this can be envisaged 
as a parallel of the sequential method underlied by passing to centre of mass frame for translations.
This analogy is in turn underlined by $P\mbox{-}Para\mbox{-}Dilatat(d)$ itself having a semidirect product structure like $Dilatat(d)$.

\m 

\n We then know from Sec I.7 that this is solved by 
\be 
\sbiQ(U, \, V)  \:=  \sbiQ \left( \,  \frac{u}{\omega} \mma \frac{v}{\omega} \, \right)                         \m .
\ee 
So far, this consists of a sequential chain rule rederivation of Sec 4. 

\m 

\n If the translational preserved equation is moreover present, this becomes 
\be 
0  \es  \left(    \left(  \frac{1}{z^2} - \frac{1}{w^2}  \right) \pa_u + 
                  \left(  \frac{1}{z^2} - \frac{1}{x^2}  \right) \pa_v + 
		          \left(  \frac{1}{z^2} - \frac{1}{y^2}  \right) \pa_{\omega}    \right) \sbiQ 
\ee 
in the special-projective characteristic coordinates.
Moreover, by tht chain rule, 
\be 
\pa_u        = U_u \pa_U + V_v \pa_V   \es  \frac{1}{\omega} \pa_U                           \m , 
\ee 
\be 
\pa_v        = U_v \pa_U + V_v \pa_V   \es  \frac{1}{\omega} \pa_V                           \m , 
\ee 
\be 
\pa_{\omega} = U_w \pa_U + V_w \pa_V   \es  - \frac{  u \, \pa_U + v \, \pa_V  }{  \omega^2  }  \m . 
\ee 
This sends our remaining PDE to 
$$
0  \es  \frac{1}{\omega}    \left( \left( \frac{1}{z^2} - \frac{1}{w^2}  -  \left( \frac{1}{z^2} - \frac{1}{y^2}  \right) U \right) \pa_U + 
                                   \left( \frac{1}{z^2} - \frac{1}{x^2}  -  \left( \frac{1}{z^2} - \frac{1}{y^2}  \right) V \right) \pa_V    \right) \sbiQ 
$$
$$
   \es  \left( \left( - \frac{1}{z} - \frac{1}{w}  + \frac{1}{z} + \frac{1}{y} \right) U \, \pa_U +
              \left( - \frac{1}{z} - \frac{1}{x}  + \frac{1}{z} + \frac{1}{y} \right) V \, \pa_V    \right) \sbiQ 
$$ 
\be 
   \es           \left( \,  ( \omega - u ) U \, \pa_U  +  ( \omega - v ) V \, \pa_V \, \right) 
    =   \omega   \left( \,  ( 1 - U )      U \, \pa_U  +  ( 1 - V ) V      \, \pa_V \, \right) \sbiQ   \m .   
\ee 
We have also used here differences of two squares and $x, y, z, w$ to $u, v, \omega$ relations in the second step, 
and $u, v, \omega$ to $U, V$ relations in the third and fourth steps. 

\m 

\n So (assuming $\omega \neq 0$, if not permute allocation of coordinates), we have the first-order homogeneous linear PDE 
\be 
\left( \, ( 1 - U ) U \, \pa_U  +  ( 1 - V ) V \, \pa_V  \, \right) \sbiQ  =  0  \m . 
\ee 
By the flow method, this is equivalent to the ODE system 
\be 
\dot{U} = ( 1 - U ) U  \m , 
\ee 
\be 
\dot{V} = ( 1 - V ) V  \m , 
\ee 
\be 
\dot{\sbiQ} = 0         \m , 
\ee  
to be treated as a Free Characteristic Problem. 
Integrating by use of partial fractions, 
\be 
t  \es  \mbox{ln} \left( \frac{U}{1 - U} \right) + \mbox{ln} \, W  \m , 
\ee 
\be 
t  \es  \mbox{ln} \left( \frac{V}{1 - V} \right)                   \m , 
\ee 
\be 
\sbiQ = \sbiQ(W)                                                     \m . 
\ee  
Eliminating $t$ between the first two of these equations yields the form for the final characteristic variable, 
\be  
W  \es  \frac{  V ( 1 - U )  }{  U ( 1 - V )  }  
   \es  \frac{  \frac{\miv}{\momega}  \left( 1 - \frac{\miu}{\momega} \right)  } { \frac{\miu}{\omega}  \left( 1 - \frac{\miv}{\momega} \right)   } 
   \es  \frac{  v ( \omega - u )}{u( \omega - v )}    
   \es  \frac{ \left( \frac{1}{\mix}  -  \frac{1}{\miz} \right)\left( \frac{1}{\miy}  - \frac{1}{\miz}  - \left( \frac{1}{\miw}  - \frac{1}{\miz} \right) \right)    }
             { \left( \frac{1}{\miw}  -  \frac{1}{\miz} \right)\left( \frac{1}{\miy}  - \frac{1}{\miz}  - \left( \frac{1}{\mix}  - \frac{1}{\miz} \right) \right)    }
   \es  \frac{ \frac{\miz - \mix}{\miz\mix} \times \frac{\miw - \miy}{\miw\miy} }
             { \frac{\miz - \miw}{\miz\miw} \times \frac{\mix - \miy}{\mix\miy} }
   \es  \frac{(z - x)(y - w)}{(z - w)(y - x)}                                                                                                                             \m : 
\ee 
the cross-ratio, now as a {\sl unique} equation to an explicit PDE.  
Thus finally we recover that 
\be 
\sbiQ  \es  \sbiQ\left( \frac{(z - x)(y - w)}{(z - w)(y - x)} \right)  \m .  
\ee

\section{Uniqueness of cross-ratios in 1-$d$. II. Arbitrary $N$}

We finally show that the previous section's method extends to arbitrary $N \geq 4$. 
We begin by solving the special-projective preserved equation piecemeal, as per Sec 3.
Next, by the chain rule, 

\n\be 
\pa_{q^i}  \es  \sum_{j = 1}^n {u^j}_{q^i} \pa_{u^j}  \es  - \frac{1}{q^{i \, 2}} \pa_{u^i}                  \m , 
\ee 

\n\be 
\pa_{q^N}  \es  \sum_{j = 1}^n {u^j}_{q^N} \pa_{u^j}  \es  - \frac{1}{q^{N \, 2}} \sum_{j = 1}^n \pa_{u^j}   \m .   
\ee 
\n The dilational preserved equation then becomes 

\n\be 
0  \es  \left(  \sum_{i = 1}^n \left( \frac{1}{q^i} - \frac{1}{q^N}   \right) \pa_{u^i}  \right) \sbiQ 
   \es          \sum_{i = 1}^n                                            u^i \pa_{u^i}          \sbiQ      \m .  
\ee 
\n We then know from Sec I.8 that this is solved by 
\be 
\sbiQ(U^{\barr})  \:=  \sbiQ \left( \frac{u^{\barr}}{u^n} \right)                                            \m , 
\ee 
for $\barr$ taking values 1 to $\bar{n} := n - 1$.  
So far, this consists of a sequential chain rule rederivation of the second half of Sec 4. 

\m 

\n If the translational preserved equation is moreover present, this gets reformulated as  
\be 
0  \es  \left( \frac{1}{q^{N \, 2}} - \frac{1}{q^{i \, 2} }  \right) \pa_{u^i} \sbiQ 
\ee 
in the special-projective characteristic coordinates.
By the chain rule, 

\n\be 
\pa_{u^{\barr}}        \es  \sum_{\bar{s} - 1}^{\bar{n}} {U^{\bar{s}}}_{U^{\barr}} \pa_u^{\bar{s}}  \es  \frac{1}{u^n} \pa_{U^{\barr}}                                            \m , 
\ee 

\n\be 
\pa_{u^n}              \es  \sum_{\bar{s} - 1}^{\bar{n}} {U^{\bar{s}}}_{U^n} \pa_u^{\bar{s}}        \es - \frac{1}{u^n}\sum_{\barr = 1}^{\bar{n}}U^{\barr}\pa_{U^{\barr}}        \m , 
\ee
our remaining PDE is sent to 

\n$$
0  \es  \frac{1}{u^n} \sum_{\barr = 1}^{\bar{n}} \left( \left( \frac{1}{q^{N \, 2}} - \frac{1}{u^{\barr \, 2}}  \right)  
                                                       \left( \frac{1}{q^{N \, 2}} - \frac{1}{q^{n \ 2}}       \right) U^{\barr}    \right)           \pa_{U^{\barr}} \sbiQ
   \es                \sum_{\barr = 1}^{\bar{n}} \left( - \frac{1}{q^{N}} - \frac{1}{u^{\barr}} +  \frac{1}{q^{N}} + \frac{1}{q^{n}} \right) U^{\barr} \pa_{U^{\barr}} \sbiQ
$$   

\n\be 
   \es                 \sum_{\barr = 1}^{\bar{n}} (  u^n - u^{\barr}  )                                                                       U^{\barr} \pa_{U^{\barr}} \sbiQ
    =   u_n            \sum_{\barr = 1}^{\bar{n}} (  1 - U^{\barr}    )                                                                       U^{\barr} \pa_{U^{\barr}} \sbiQ  \m ,   
\ee
using analogous moves to those declared in the $N = 4$ version.  

\m 

\n So (assuming $u^n \neq 0$, if not permute allocation of coordinates), we have the first-order homogeneous linear PDE 

\n\be 
\sum_{\barr = 1}^{\bar{n}} ( 1 - U^{\barr})U\pa_{U^{\barr}} \sbiQ  \es  0  \m . 
\ee 
By the flow method, this is equivalent to the ODE system 
\be 
\dot{U}^{\barr} = (  1 - U^{\barr}  )U^{\barr}                            \m , 
\ee 
\be 
\dot{\sbiQ} = 0                                                            \m , 
\ee  
to be treated as a Free Characteristic Problem. 
Integrating by use of partial fractions, 
\be 
t  \es  \mbox{ln}  \left(  \frac{  U^{\widetilde{r}}  }{  1 - U^{\widetilde{r}}  }  \right) + \mbox{ln} \, W^{\widetilde{r}}      \m , 
\label{tilder}
\ee 
\be 
t  \es  \mbox{ln}  \left(  \frac{  U^{\bar{n}}  }{  1 - U^{\bar{n}}  }  \right)                                                   \m , 
\label{teq}
\ee 
\be 
\sbiQ = \sbiQ(W^{\widetilde{r}})                                                                                                    \m . 
\ee  
Eliminating $t$ from (\ref{teq}) in (\ref{tilder}) yields the form for the final characteristic variables, 
\be  
W^{\widetilde{r}}  \es  \frac{U^{\bar{n}}(1 - U^{\widetilde{r}})}{U^{\widetilde{r}}(1 - V^{\bar{n}})}  
                   \es  \frac{u^{\bar{n}}(u^n - u^{\widetilde{r}})}{u^{\widetilde{r}}(u^n - u^{\bar{n}})}
                   \es  \frac{(q^N - q^{\bar{n}})(q^n - q^{\widetilde{r}})}{(q^n - q^{\bar{n}})(q^N - q^{\widetilde{r}} )}        \m : 
\ee 
the cross-ratio, now as a {\sl unique} equation to an explicit PDE.  
Thus finally we recover that 
\be 
\sbiQ  \es  \sbiQ\left( \frac{(q^N - q^{\bar{n}})(q^n - q^{\widetilde{r}})}{(q^n - q^{\bar{n}})(q^N - q^{\widetilde{r}} )} \right)  
      \es  \sbiQ( \, {\bm{;}} \, )                                                                                                   \m :   
\ee
suitably-smooth functions of cross-ratios.  

\m 

\n Note that the cross-ratios arise here in the form of a basis made by using 3 points $N$, $n$, $\bar{n}$ as fixed reference points 
with respect to which to form the cross-ratio with the $\widetilde{n} := N - 3$ remaining points $\widetilde{r}$. 
Because of this, moreover, it is clear that $N \geq 4$ is required. 

\section{Conclusion}
%
{            \begin{figure}[!ht]
\centering
\includegraphics[width=0.85\textwidth]{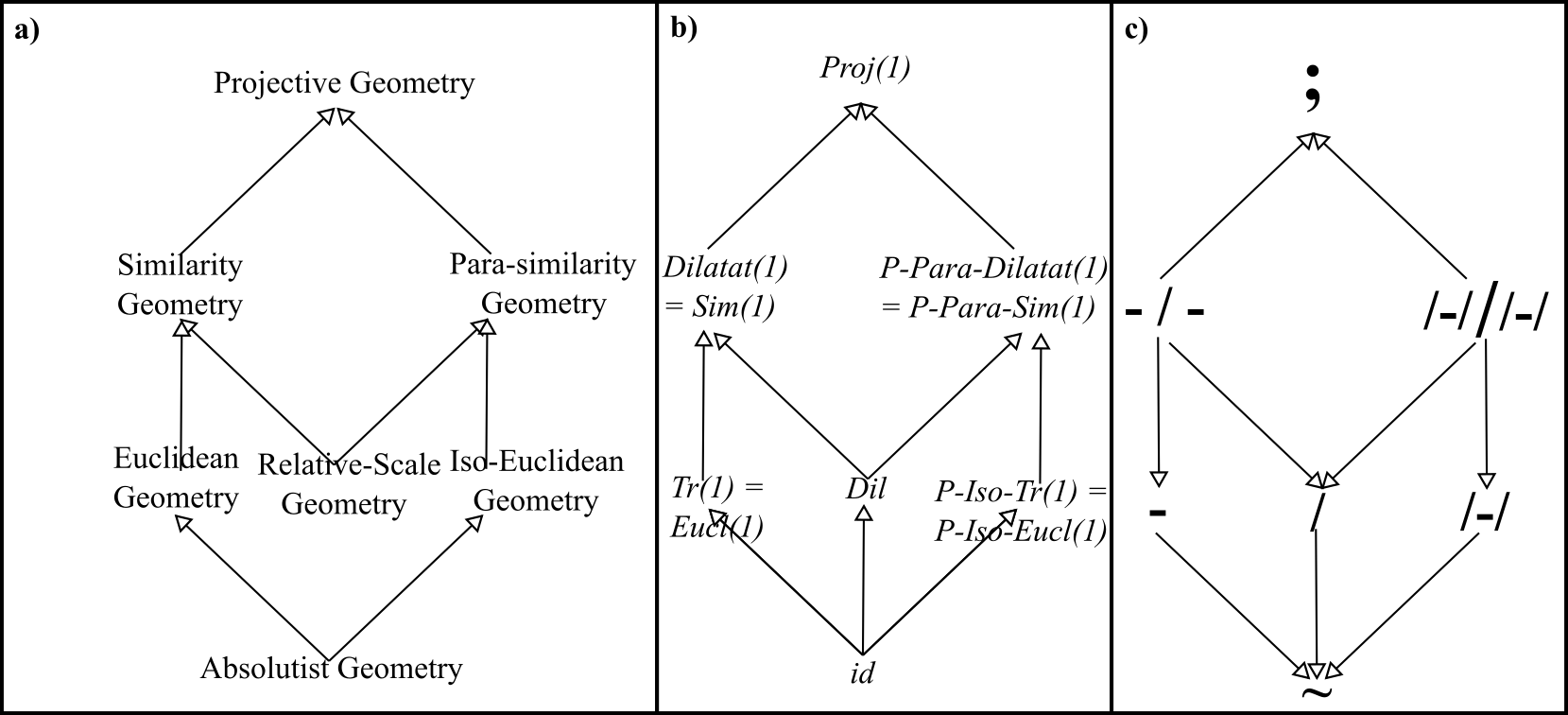}
\caption[Text der im Bilderverzeichnis auftaucht]{        \footnotesize{Lattices of a) 1-$d$ notions of geometry, 
                                                                                    b) their corresponding automorphism groups, 
																			    and c) the corresponding dual lattice of preserved quantities. }}
\label{Proj-1-Latt} \end{figure}          }

\n We found preserved quantities for the $P$-$Iso$-$Tr(1)$ = $P$-$Iso$-$Eucl(1)$ Geometry whose automorphisms consist solely of special-projective transformations $Q$. 
These are suitably smooth functions of differences of reciprocals. 
Thus they do not coincide with the mere differences of $Tr(1) = Eucl(1)$ despite $Eucl(1) \cong P\mbox{-}Iso\mbox{-}Eucl(1)$, 
the difference in representation between $Q$ and the translational generator $P$ sufficing to have this effect.  

\m 

\n Upon including dilations $D$ as well, we found preserved quantities for the corresponding 

\n $P$-$Para$-$Dilatat(1)$ = $P$-$Para$-$Sim(1)$ = $P$-$Para$-$Aff(1)$ Geometry.
These are suitably smooth functions of ratios differences of reciprocals. 
Thus they do not coincide with the mere ratios differences of 

\n $Dilatat(1) = Sim(1) = Aff(1)$, despite the algebra for this differing by a single sign from our case. 
(This sign difference also accounts for us calling these `Para' rather than `Iso' Geometries).

\m 

\n We finally derived that 1-$d$ projective preserved quantities are suitably smooth functions of cross-ratios in Secs 5 and 6, 
establishing these to be moreover the {\sl unique} functional form solving the 1-$d$ projective preserved equations system's Free Characteristic Problem in Secs 7 and 8.   
The current Article's analysis points moreover to a new interpretation of cross-ratio.
Namely that cross-ratio functional dependence is that functional dependence which is concurrently of ratios of differences and of differences of reciprocals, 
as can be read off Sec 6's compatibility equation.
This is a significant result firstly due to the importance of cross-ratios in Projective Geometry and secondly because of Application 2 below. 
 
\m 

\n{\bf Application 1} The above analysis featuring novel {\it partially} projective preserved quantities serves to disqualify the `counterexamples' to lattice duality of \cite{AObs3}, 
as these `counterexamples' failed to factor in the possibility of partially conformal preserved quantities (themselves covered in Article V). 
Now that preserved quantities are viewed as intersections of characteristic surfaces, it has become clear that the lattice of preserved quantities is dual to that of 
sums-over-points of generators, and, by extension via Article I's Bridge Theorem, that the lattice of observables is dual to that of first-class constraints.
We can thus re-issue \cite{AObs3} free from this lacuna. 

\m 

\n{\bf Application 2} The current Article is moreover a useful prototype as regards systematically solving PDEs to obtain the more involved 
higher-$d$ projective preserved quantities in Article IV, alongside yet further partially projective preserved quantities from interplay with rotations and affine transformations. 

\m 

\n{\bf Acknowledgments} I thank Chris Isham and Don Page for previous discussions.  
Reza Tavakol, Malcolm MacCallum, Enrique Alvarez and Jeremy Butterfield for support with my career.


\end{document}